\documentclass[12pt]{article}

\usepackage[fleqn]{amsmath}
\usepackage{amsfonts}
\usepackage{amssymb}
\usepackage{epsfig}
\usepackage{geometry}
\usepackage{ctable}
\usepackage[all,cmtip]{xy}
\usepackage{graphicx}
\usepackage{float}

\newcommand{\drm}{\mathrm{d}}

\newcommand{\Ddt}{\frac{\drm\phantom{s}}{\drm t}}

\newcommand{\pddt}{\frac{\partial\phantom{t}}{\partial t}}

\newcommand{\tpddt}{{\textstyle{\frac{\partial\phantom{t}}{\partial t}}}}

\newcommand{\refeq}[1]{(\ref{#1})}

\newcommand{\mbare}{m_{\text{b}}}
\newcommand{\mfield}{m_{\text{f}}}

\newcommand{\vect}[1] {\boldsymbol{{ #1}} }

\newcommand{\qv}[1]{{\textbf{\textsf{#1}}}}

\newcommand{\Rset}{\mathbb{R}}

\newcommand{\ID}{{\boldsymbol{I}_{3\times3}^{}}} 

\newcommand{\aV}{\vect{a}}              

\newcommand{\fV}{\vect{f}}              
\newcommand{\FV}{\vect{F}}              
\newcommand{\nV}{{\vect{n}}}		
\newcommand{\pV}{{\vect{p}}}            
\newcommand{\qV}{{\vect{q}}}            
\newcommand{\sV}{{\vect{s}}}            
\newcommand{\vV}{{\vect{v}}}            

\newcommand{\DLWr}{{\DV}_{\text{\textsc{lw}}}^{\text{\tiny{ret}}}} 
\newcommand{\HLWr}{{\HV}_{\text{\textsc{lw}}}^{\text{\tiny{ret}}}} 

\newcommand{\NullV}{\vect{0}}

\newcommand{\BV}{\pmb{{\cal B}}}

\newcommand{\DV}{\pmb{{\cal D}}}
\newcommand{\EV}{\pmb{{\cal E}}}

\newcommand{\HV}{\pmb{{\cal H}}}

\newcommand{\PiV}{\boldsymbol{\Pi}}
\newcommand{\nab}{\vect{\nabla}}

\renewcommand{\leq}{\leqslant}
\renewcommand{\geq}{\geqslant}

\newcommand{\crprd}{{\boldsymbol\times}}

\newcommand{\ee}{\mathit{e}}
\newcommand{\eEL}{\mathit{e}}

\reversemarginpar

\numberwithin{equation}{section}

\begin{document}

       
\title{\sc{a study of the radiation-reaction on} \\ \sc{a point charge that moves along a constant applied electric field 
in an electromagnetic
 \sc{b}{\tiny\sc{opp}}-\sc{l}{\tiny\sc{and\'e}}-\sc{t}{\tiny\sc{homas}}-\sc{p}{\tiny\sc{odolsky}} vacuum}}

\author{\textbf{H. K. Carley$^1$ and M. K.-H. Kiessling$^2$}\\
\small           $^1$ Department of Mathematics, \\
\small           New York City College of Technology, CUNY\\
\small           300 Jay Street, Brooklyn, New York, NY 11201, USA\\
\small           $^2$ Department of Mathematics, \\ 
\small           Rutgers, The State University of New Jersey,\\
\small           110 Frelinghuysen Rd., Piscataway, NJ 08854, USA\\ 
\textrm{\tiny Version of March 13, 2023. Typeset with \LaTeX\ on: }}
\maketitle

\thispagestyle{empty}
\vspace{-1truecm}
\begin{abstract}
\noindent 
 The relativistic problem of motion of a classical electrical point charge that has been placed
between the plates of a charged capacitor and then released from rest
is well-posed in Bopp--Land\'e--Thomas--Podolsky (BLTP) electrodynamics.
 That theory introduces a single new parameter, Bopp's $\varkappa$, a reciprocal length.
 The present article concerns the small-$\varkappa$ regime. 
 Radiation-reaction effects on the motion are shown to appear at order $\varkappa^3$.
 It is found that in the initial phase the motion is accurately accounted for by test particle theory, with 
the inertia determined by the bare mass of the particle.
 Subsequently, radiation-reaction effects cause substantial deviations from the test particle motion.
\end{abstract}

\bigskip
\centerline{IN REMEMBRANCE OF}
\centerline{\bf Detlef D\"urr}
\bigskip

\vfill
\hrule
\smallskip

\copyright(2023) \small{The authors. Reproduction of this preprint, in its entirety, is permitted

\hspace{1.5truecm} for non-commercial purposes only.}

\newpage


                \section{Introduction}
\noindent

 In this tribute to Detlef D\"urr we focus on a lesser known scientific passion of Detlef, 
i.e.,  ``lesser known in comparison to Bohmian Mechanics,''
and this is the classical electromagnetic radiation-reaction problem. 
 Here is how Detlef once characterized the situation:
\begin{quote}
When people realized that there is a problem, quantum physics was invented. 
 Then everyone began to work on quantum physics and eventually the problem was forgotten. 
 But it still exists. (Detlef D\"urr, private communication to the senior author, sometime
in the mid 1980s.)
\end{quote}
 The problem, in a nutshell, is this:
The symbolic system of equations of Lorentz electrodynamics with point charges is notoriously ill-defined.
 The energy and momentum densities of the electromagnetic Maxwell--Lorentz fields of a point charge source are not 
integrable over any neighborhood of the point charge, and also the Lorentz formula for the electromagnetic ``self''-force 
on such a point charge source is ill-defined. 
 More recently \cite{DeckertHartenstein} it was noted that also the Lorentz formula for the electromagnetic force
of one point charge source onto another becomes ill-defined after relatively short times.

\hspace{-8pt}
 Undeterred by infinities \cite{PoissonETal}, physicists have tried to extract the force of radiation-reaction 
on an accelerated point charge by analyzing the power emitted by it towards $|\sV| = \infty$, per
the retarded Li\'enard--Wiechert fields.
 Von Laue \cite{Laue} obtained the expression $\frac{2e^2}{3c^3}P_{\qv{u}(\tau)}^\perp\cdot\frac{d^2}{d\tau^2}\qv{u}(\tau)$ for the 
Minkowski force four-vector due to the radiation (cf. \cite{JacksonBOOKb}); here, $\tau$ denotes proper time and
$P_{\qv{u}(\tau)}^\perp\cdot$ the four-projection onto the subspace that is four-orthogonal to the four-velocity 
$\qv{u}(\tau)$ ($=\frac{d}{d\tau}\qv{q}(\tau)$).
 Its third derivative of the particle's spacetime location $\qv{q}(\tau)$ has been the cause of consternation.
 In particular, it vanishes during intervals of constant four-acceleration when the Larmor formula predicts radiation.
 Moreover, when non-zero, it changes the initial-value problem from second to third order, and almost all solutions then
display unphysical behavior.
 It has been argued that a way out of this ``third-order'' dilemma is offered by the fact that the radiation-reaction on the particle 
should cause only a small correction term to the test particle type force. 
 The co-variant version of the equations of test particle motion reads
$\frac{d^2}{d\tau^2}\qv{q}(\tau) = \frac{e}{mc} \qv{F}(\qv{q}(\tau))\cdot\qv{u}(\tau)$,
where $\qv{F}(\qv{s})$ is the Faraday tensor of the applied fields at a spacetime point $\qv{s}$.
 Hence the expression $\frac{d^2}{d\tau^2}\qv{u}(\tau)$ in von Laue's four vector should be
interpreted as stand-in for the first-order proper time derivative of $\frac{e}{mc}\qv{F}(\qv{q}(\tau))\cdot\qv{u}(\tau)$,
which will not have derivatives of $\qv{q}(\tau)$ higher than second order. 
 The so radiation-reaction-corrected equation of motion,
sometimes called the Eliezer--Ford--O'Connell (EFC) equation of motion, reads
\begin{equation}
\label{eq:EFOCeOFmotion}
\frac{d^2}{d\tau^2}\qv{q}(\tau) = \frac{e}{mc}\qv{F}(\qv{q}(\tau))\cdot\qv{u}(\tau)
+  
\frac23\frac{e^3}{mc^4}P_{\qv{u}(\tau)}^\perp \cdot\frac{d}{d\tau}\big(\qv{F}(\qv{q}(\tau))\cdot\qv{u}(\tau)\big).
\end{equation}
 Landau and Lifshitz approximated (\ref{eq:EFOCeOFmotion}) further by replacing all first-order 
proper time derivatives of $\qv{u}(\tau)$ obtained from $\frac{d}{d\tau}\big(\qv{F}(\qv{q}(\tau))\cdot\qv{u}(\tau)\big)$ 
by $\frac{e}{mc} \qv{F}(\qv{q}(\tau))\cdot\qv{u}(\tau)$.
 This approximation to (\ref{eq:EFOCeOFmotion}) is known as the Landau--Lifshitz (LL) equation of motion.

 Equation (\ref{eq:EFOCeOFmotion}), and also its Landau--Lifshitz approximation, enjoy some practical
successes.
 Interestingly, this practical success story has a serious blemish.
 Namely, for a point charge that moves along a constant applied electrostatic field
the LL equation of motion simply reproduces the test particle motion, for its 
radiation-reaction force term vanishes in this situation; cf. \cite{PMD}. 
 One may hope to obtain the radiation-reaction effects on a point charge that moves along a constant
electric field by pushing the expansion of the EFC equation further until a non-vanishing radiation-reaction 
force term is obtained.
 However, all higher-order terms obtained from such an expansion vanish also.

 As an aside, we mention that the LL equation has been derived rigorously as an effective equation of motion
in a spacetime adiabatic limit,
not for a point charge but for an extended charge distribution's 
geometric center, and with $m$ standing for $\mbare+\mfield$, where $\mbare$ is a bare mass and $\mfield$ a 
field energy contribution; see \cite{Spohn}.
 Thus, if one works with extended charge distributions, as in the Abraham--Lorentz-type classical electron theory
\cite{LorentzENCYCLOP},
and endows the particle with a non-vanishing $\mbare$, then one may realistically hope to obtain 
higher-order radiation-reaction corrections that do not vanish for motion along a constant electric field.
 The LL equation of motion was also obtained in an asymptotic expansion about a ``vanishing particle limit,''
which captures the motion of a particle with extended charge and mass distributions
in the limit of vanishing size, mass, and charge, yet with nonzero charge-to-mass ratio, see \cite{GHW}.
 However, higher-order terms in such expansions will eventually depend on largely arbitrary assumptions 
about the structure of the extended distributions.
 Moreover, the formulation of a properly Lorentz co-variant model with extended charged particles
\cite{AppKieAOP} involves a non-trivial foliation of Minkowski spacetime and poses 
conceptual challenges for the initial value problem.

 In this paper we are interested in the classical theory of motion for a true point charge that interacts with the
electromagnetic fields it generates.
 In \cite{KiePRD} the first {well-posedness} result of the joint initial value problem for the 
evolution of the electromagnetic fields and the relativistic motion of $N$ point charges was announced,
not for the ill-defined Lorentz electrodynamics with point charges, but for an electrodynamic model that goes back to work by Bopp
\cite{BoppA,BoppB}, Land\'e--Thomas \cite{Lande}, \cite{LandeThomas}, and Podolsky \cite{Podolsky} (BLTP).
 The BLTP model replaces Maxwell's law of the electromagnetic vacuum 
(viz. $\HV=\BV$ and $\EV=\DV$) with a linear differential relation. 
 A well-defined equation of motion was supplied in \cite{KiePRD}.
 The proof of well-posedness of the joint Cauchy problem will be published in \cite{KTZonBLTP}.
 Also the scattering problem for a single particle that encounters a localized potential is well-posed \cite{VuMaria}.
 These authors showed that in this problem the ``self''-force formula of \cite{KiePRD} can be converted into a formal 
Lorentz-type expression that involves integration over the whole past of the particle motion, first proposed in
\cite{LandeThomas} and further studied in \cite{Zayats} and \cite{GratusETal}.

In the following we demonstrate that BLTP electrodynamics captures the radiation-reaction on a point charge that
is released from rest in a constant applied electric field. 
 We understand our work as part of a proof-of-concept.
 In principle our approach can handle also more realistic models than BLTP electrodynamics, see \cite{KiePRD}.


     \section{\hspace{-10pt}BLTP electrodynamics with a single point charge} 

 The electromagnetic vacuum in BLTP electrodynamics is defined by the two equations
\begin{alignat}{1}
        \HV(t,\sV)  
& = \label{eq:BLTPlawBandH}
       \left(1  + \varkappa^{-2}\square\,\right) \BV(t,\sV) \, , \\ 
        \DV(t,\sV) 
& =
        \left(1  + \varkappa^{-2}\square\,\right) \EV(t,\sV) \, ;
\label{eq:BLTPlawEandD}
\end{alignat}
in \refeq{eq:BLTPlawBandH} and \refeq{eq:BLTPlawEandD}, the parameter $\varkappa^{-1}$ is the ``Bopp length'' \cite{BoppA,BoppB},
and $\square \equiv c^{-2}\partial_t^2 -\Delta$ is the d'Alembertian, with $c$ the vacuum speed of light.
 The evaluations $\HV(t,\sV)$, $\BV(t,\sV)$, $\EV(t,\sV)$, and $\DV(t,\sV)$ of the fields at the space point
$\sV\in\Rset^3$ and instant of time $t\in\Rset$ are defined in any convenient flat foliation of Minkowski spacetime 
into space \&\ time. 
 These fields satisfy the familiar system of pre-metric Maxwell field equations, which consist of two evolution equations
\begin{alignat}{1}
\textstyle
\pddt{\BV(t,\sV)}
&= \label{eq:MdotB}
        - c \nab\crprd\EV(t,\sV) \, ,
\\
\textstyle
\pddt{\DV(t,\sV)}
&= 
        + c\nab\crprd\HV(t,\sV)  - 4\pi \ee \delta_{\qV(t)}(\sV){\vV}(t)\, ,
\label{eq:MdotD}
\end{alignat}
and two constraint equations
\begin{alignat}{1}
        \nab\cdot \BV(t,\sV)  
&= \label{eq:MdivB}
        0\, ,
\\
        \nab\cdot\DV(t,\sV)  
&=
        4 \pi \ee \delta_{\qV(t)}(\sV)\, .
\label{eq:MdivD}
\end{alignat}
 Here,  $\ee (>0)$ is the elementary electric charge,
$\qV(t)\in\Rset^3$ its position and $\vV(t) \in\Rset^3$ its velocity at time $t$.

  The particle's velocity is defined as usual to be the time-derivative of its position vector,
\begin{equation}
\Ddt \qV(t)
= : \label{eq:dotQisV}
\vV(t).
\end{equation}
   In the relativistic generalization of Newton's point mechanics by Einstein, Lorentz, and Poincar\'e, 
the velocity $\vV(t)$, in turn, changes with time according to
\begin{equation}
\Ddt   \frac{\vV(t)}{\sqrt{1 - \frac{1}{c^2}|{\vV}(t)|^2}}
= \label{eq:EinsteinNewtonEQofMOT}
\frac{1}{\mbare}\fV(t);
\end{equation}
here, $\mbare \neq 0$ is the \emph{bare inertial rest mass} of the particle, 
and $\fV(t)$ is the total electromagnetic force acting on it.
 Following Poincar\'e (cf. \cite{MillerBOOK}) we~define~it~as  (cf. \cite{KiePRD})
\begin{equation}
\fV(t) := \label{eq:POINCAREforce}
\eEL\,\EV^{\mbox{\tiny{hom}}}
 - \Ddt \int_{\Rset^3} \PiV^{\mbox{\tiny{field}}}(t,\sV) \drm^3s,
\end{equation}
where  $\EV^{\mbox{\tiny{hom}}}$ is a constant applied electric field (an idealization of the field between the plates of a 
capacitor), and $\PiV^{\mbox{\tiny{field}}}(t,\sV)$ is the momentum vector-density of the Maxwell-BLTP fields 
\begin{equation}
\textstyle
4\pi c \PiV^{\mbox{\tiny{field}}}
= \label{eq:PiMBLTP}
\DV\crprd\BV + \EV\crprd\HV - \EV\crprd\BV 
- \varkappa^{-2} \big(\nabla\cdot\EV\big)\Big(\nabla\crprd\BV - \frac{1}{c}\pddt\EV\Big).
\end{equation}

     \section{\hspace{-10pt}The initial data} 

 As announced in \cite{KiePRD} and shown in \cite{KTZonBLTP}, BLTP electrodynamics is well-posed as a joint
initial value problem for the fields and the point charge, requiring initial data 
$\BV(0,\sV)$, $\DV(0,\sV)$, $\EV(0,\sV)$, $(\pddt\EV)(0,\sV)$ for the fields, and $\qV(0)$ and $\vV(0)$ for
the particle. 
 The data for $\BV$ and $\DV$ are constrained by the divergence equations, and  $\vV(0)$ by $|\vV(0)|<c$.

 In the ensuing sections we discuss these BLTP-dynamical equations for a single point charge moving along
the constant applied electric field $\EV^{\mbox{\tiny{hom}}}$, starting from rest,
with the initial fields the sum of the external field and the electrostatic field of the point charge.
 Thus, for the particle we have
\begin{equation}
\qV(0)=\NullV
\qquad \mbox{and} \qquad
\vV(0)=\NullV.
\end{equation}
 For the fields we have
\begin{equation}\label{Dinit}
\DV(0,\sV)\equiv \EV^{\mbox{\tiny{hom}}} +\ee \frac{\sV}{|\sV|^3}
\end{equation}
and
\begin{equation}\label{Einit}
\EV(0,\sV)
\equiv \EV^{\mbox{\tiny{hom}}} + \ee \frac{1 - (1+\varkappa |\sV|)e^{-\varkappa |\sV|}}{|\sV|^2}\frac{\sV}{|\sV|}.
\end{equation}
 We also have $\big(\partial_t \EV\big)(0,\sV)\equiv \NullV$, as well as $\BV(0,\sV) \equiv \NullV.$

     \section{\hspace{-10pt}The solution of the field equations} 

 For the initial data of our problem the electromagnetic fields outside the forward light cone of 
the initial location of the particle remain precisely the electrostatic fields, i.e., the magnetic field $\HV$ 
and the magnetic induction field $\BV$ vanish, while the electric displacement field $\DV(t,\sV)$ is given by \refeq{Dinit}
and the electric field $\EV(t,\sV)$ is given by \refeq{Einit}, for all $t\geq 0$.

 Inside the forward light cone of the initial particle location, but away from the particle position at $t$,
the fields $\DV(t,\sV)$ and $\HV(t,\sV)$ are for all $t\geq 0$ given by
$\DV = \EV^{\mbox{\tiny{hom}}} + \DLWr$ \&\ $\HV =\HLWr$, with 
(the acceleration vector of the point charge is highlighted in {\color{red}red})
\begin{alignat}{1}
\hskip-.6truecm
{\DLWr(t,\sV)} &=\label{eq:LWsolD}
  \ee \frac{c^2-|\vV|^2}{|\sV-\qV|^2} 
\frac{{c\nV(\qV,\sV)}_{\phantom{!\!}}-{\vV}}{\bigl(\textstyle{c-\nV(\qV,\sV)\cdot {\vV}}\bigr)^{\!3}} 
\Biggl.\Biggr|_{\mathrm{ret}}
+\ee 
\frac{\nV(\qV,\sV)\crprd
\bigl[\bigl(c\nV(\qV,\sV)_{\phantom{!\!}}-{\vV}\bigr)\crprd{\color{red}\aV}\bigr]}{{|\sV-\qV|}
\bigl(\textstyle{c-\nV(\qV,\sV)\cdot {\vV}}\bigr)^{\!3}}
\Biggl.\Biggr|_{\mathrm{ret}},\hskip-1truecm
\\
\hskip-1truecm
{\HLWr(t,\sV)} 
&= \label{eq:LWsolH}
        \nV(\qV,\sV)|_{_{\mathrm{ret}}}\crprd {\DLWr(t,\sV)}
\, \vspace{-10pt}
\end{alignat}
the retarded Li\'enard--Wiechert fields. 
 Here, $\nV(\qV,\sV) = \frac{\sV-\qV}{|\sV-\qV|}$ is a \emph{normalized} vector from $\qV$ to $\sV$,
 the notation ``$|_{\mathrm{ret}}$'' means that $(\qV,\vV,{\color{red}\aV})= (\qV,\vV,{\color{red}\aV})(t^{\mathrm{ret}})$ 
to the left of ``$|_{\mathrm{ret}}$,''
with $t^{\mathrm{ret}}(t,\sV)$ being defined implicitly by $c(t-t^{\mathrm{ret}}) = |\sV-\qV(t^{\mathrm{ret}})|$;
inside the initial forward light cone, $0< t^{\mathrm{ret}} <t$. 
The terms $\propto{\color{red}\aV}$ in (\ref{eq:LWsolD}) and (\ref{eq:LWsolH}) account for the radiation.

 Note that the electromagnetic Li\'enard--Wiechert fields $\HLWr$ and $\DLWr$ exhibit both
a $\propto 1/r^2$ singularity and a $\propto 1/r$ singularity, where $r$ denotes $|\sV-\qV(t)|$; they each
have a directional singularity at the location of the  point charge source, too.

 Similarly, inside --- and on --- the forward light cone of the initial particle location, but away from the particle position at $t$,
the MBLTP field solutions $\BV(t,\sV)$ and $\EV(t,\sV)$ for $t\geq 0$ are given by 
$\BV=\BV_0+ \BV_1$ and $\EV = \EV_0 + \EV_1$, with $\BV_0\equiv\NullV$ and $\EV_0\equiv \EV^{\mbox{\tiny{hom}}}$, and
\begin{alignat}{1}
\hspace{-20pt}
{\EV_1(t,\sV)} =\; \nonumber
 & \ee \varkappa^2\Bigl(\frac{1 - (1+\varkappa |\sV|)e^{-\varkappa |\sV|}}{\varkappa^2|\sV|^2} -\tfrac12 \Bigr)\frac{\sV}{|\sV|} + 
\ee \varkappa^2 \int_{0}^{ct-|\sV|}
\tfrac{J_2\!\bigl(\varkappa\sqrt{c^2(t-t')^2-|\sV|^2}\bigr)}{{c^2(t-t')^2-|\sV|^2}^{\phantom{n}} }
 \sV \drm{(ct')} +
\\  \label{eq:EjsolMBLTP} 
 & \ee \varkappa^2\tfrac12\tfrac{\nV(\qV^{},\sV)-{\vV^{}}/{c}}{1-\nV(\qV^{},\sV)\cdot{\vV}/{c}}
\Big|_{\mathrm{ret}} \, - \\ \nonumber
 &
\ee \varkappa^2 \int_{0}^{t^\mathrm{ret}(t,\sV)}
\tfrac{J_2\!\bigl(\varkappa\sqrt{c^2(t-t')^2-|\sV-\qV(t')|^2}\bigr)}{{c^2(t-t')^2-|\sV-\qV(t')|^2}^{\phantom{n}} }
\left(\sV-\qV^{}(t') - \vV^{}(t')(t-t')\right)c\drm{t'} , \\ 
\hspace{-20pt}
{\BV_1(t,\sV)} =\; \label{eq:BjsolMBLTP}
&\ee \varkappa^2 \tfrac{1}{2} 
\tfrac{{\color{black}\vV^{}\crprd\nV(\qV^{},\sV)/c}}{1-\nV(\qV^{},\sV)\cdot{\vV}/{c}}
\Big|_{\mathrm{ret}}
\, -  \\ \nonumber
 &
 \ee \varkappa^2 \int_{0}^{t^\mathrm{ret}(t,\sV)}
\tfrac{J_2\!\bigl(\varkappa\sqrt{c^2(t-t')^2-|\sV-\qV(t')|^2 }\bigr)}{{c^2(t-t')^2-|\sV-\qV(t')|^2}^{\phantom{n}} }
{\vV^{}(t')}\crprd \left(\sV-\qV^{}(t') 
\right)\drm{t'} .
\end{alignat}
 The fields $\BV(t,\sV)$ and $\EV(t,\sV)$ 
are globally bounded in $\sV$ for each $t$, and away from the point charge they are Lipschitz-continuous in $\sV$, including
across the initial forward light cone. 

 Similarly, 
\begin{alignat}{1}
\qquad\qquad\nabla\cdot\EV(t,\sV) =\; \nonumber
 & \ee \varkappa^2 \frac{e^{-\varkappa |\sV|} - 1}{|\sV|}
+ 
\ee \varkappa^3 \int_{0}^{ct-|\sV|}
\tfrac{J_1\!\bigl(\varkappa\sqrt{c^2(t-t')^2-|\sV|^2}\bigr)}{\sqrt{c^2(t-t')^2-|\sV|^2}^{\phantom{n}} }
 \drm{(ct')} +
\\  \label{eq:LWsolPHI}
& \ee^{}\varkappa^2 
\tfrac{1}{\bigl(1-\nV(\qV,\sV)\cdot {\vV^{}}/{c}\bigr)}
\tfrac{1}{|\sV-\qV|}
\Bigl.\Bigr|_{\mathrm{ret}}
- \\ \nonumber
&
\ee^{} \varkappa^3 \int_{0}^{t^\mathrm{ret}(t,\sV)}
\tfrac{J_1\!\bigl(\varkappa\sqrt{c^2(t-t')^2-|\sV-\qV(t')|^2}\bigr)}{\sqrt{c^2(t-t')^2-|\sV-\qV(t')|^2}^{\phantom{n}} }
c \drm{t'} ,
\end{alignat}
and
\begin{alignat}{1}
\big(\nabla\crprd\BV - {\textstyle{\frac{1}{c}\pddt}}\EV\big)(t,\sV) 
=\; \label{eq:LWsolA}
& \ee^{}\varkappa^2\tfrac{1}{\bigl(1-\nV(\qV,\sV)\cdot {\vV^{}}/{c}\bigr)}
\tfrac{1}{|\sV-\qV|}\tfrac{\vV^{}}{c}
\Bigl.\Bigr|_{\mathrm{ret}}
- \\ \nonumber
 & \ee^{} \varkappa^3  \int_{0}^{t^\mathrm{ret}(t,\sV)} 
\tfrac{J_1\!\bigl(\varkappa\sqrt{c^2(t-t')^2-|\sV-\qV(t')|^2}\bigr)}{\sqrt{c^2(t-t')^2-|\sV-\qV(t')|^2}^{\phantom{n}} }
\vV^{}(t')\drm{t'} .
\end{alignat}

     \section{\hspace{-10pt}Evaluation of the radiation-reaction force} 
 With the help of these solution formulas, the electromagnetic force of the MBLTP field on its point charge source
can be computed as follows. 
 Since each electromagnetic field component is the sum of a vacuum field and a sourced field, the bilinear 
$\PiV^{\mbox{\tiny{field}}}$
decomposes into a sum of three types of terms: the vacuum-vacuum terms, the source-source terms, and the mixed vacuum-source terms.
 In our problem the vacuum field is $\EV^{\mbox{\tiny{hom}}}$; it does not contribute to $\PiV^{\mbox{\tiny{field}}}$,
but appears separately at rhs\refeq{eq:POINCAREforce}.
 As exlained in \cite{KiePRD}, this term is not put in by hand but is a contribution to the momentum 
balance due to a surface integral at ``$|\sV|=\infty$.''
 Hence the only contribution to rhs\refeq{eq:POINCAREforce} from $\PiV^{\mbox{\tiny{field}}}$ is the 
source-source contribution, a ``self''-field force in BLTP electrodynamics. 
 Thus, \refeq{eq:POINCAREforce} is given by 
\begin{equation}\label{eq:totalF}
\fV(t) 
=
 \ee \EV^{\mbox{\tiny{hom}}}
+
\fV^{\mbox{\tiny{self}}}[\qV,\vV;{\color{red}\aV}](t),
\end{equation} 
where $\ee \EV^{\mbox{\tiny{hom}}}$ is the Lorentz force 
evaluated with the vacuum field (i.e., a ``test particle contribution'' to the total force), and
(after taking advantage of hyperbolicity; cf. \cite{KiePRD})
\begin{alignat}{2}\label{eq:selfFa}
\hspace{-20pt}
\fV^{\mbox{\tiny{self}}}[\qV,\vV;{\color{red}\aV}](t)
 \equiv &  - \frac{\drm}{\drm{t}} \displaystyle\int_{B_{ct}(\qV_0)}
\Bigl( \PiV^{\mbox{\tiny{field}}}_{\mbox{\tiny{source}}}(t,\sV) - 
\PiV^{\mbox{\tiny{field}}}_{\mbox{\tiny{source}}}(0,\sV-\qV_0-\vV_{\!0}t)\Bigr) d^3{s}  \\
 =& - \frac{\drm}{\drm{t}} \displaystyle\int_{B_{ct}(\qV_0)}
 \PiV^{\mbox{\tiny{field}}}_{\mbox{\tiny{source}}}(t,\sV)  d^3{s}  ,\label{selfFb}
\end{alignat}
with $\PiV^{\mbox{\tiny{field}}}_{\mbox{\tiny{source}}}$
given by \refeq{eq:PiMBLTP} with $(\BV,\DV-\EV^{\mbox{\tiny{hom}}},\EV-\EV^{\mbox{\tiny{hom}}},\HV)$ in place of $(\BV,\DV,\EV,\HV)$.
 To go from \refeq{eq:selfFa} to 
 \refeq{selfFb} we made use of the initial data $\qV_0=\NullV$ and $\vV_0=\NullV$, and 
$\PiV^{\mbox{\tiny{field}}}_{\mbox{\tiny{source}}}(0,\sV)\equiv \NullV$.

 The ``self''-field force can be evaluated using retarded spherical coordinates $(r,\vartheta,\varphi)$ to carry out the
$\drm^3{s}$ integrations over the ball ${B_{ct}(\qV_0)}$, after which one can differentiate w.r.t. $t$.
 For this very special problem of straight line motion of a charge starting from rest at the origin, 
this yields
\begin{alignat}{2}\label{eq:selfFexpl}
\hspace{-20pt}
\fV^{\mbox{\tiny{self}}}[\qV,\vV;{\color{red}\aV}](t)
&=  \frac{\ee^2} {4\pi } \biggl[ \biggr.
 - {\mathbf{Z}}_{\boldsymbol{\xi}}^{[2]}(t,t) 
\\ \notag
& \qquad\quad  -\!\!\! \;{\textstyle\sum\limits_{0\leq k\leq 1}}\! c^{2-k}(2-k)\!\!
\displaystyle  \int_0^{t}\!  
{\mathbf{Z}}_{\boldsymbol{\xi}}^{[k]}\big(t,t^{\mathrm{r}}\big)
(t- t^{\mbox{\tiny{r}}})^{1-k} \drm{t^{\mbox{\tiny{r}}}} 
\\ \notag
& \qquad\quad -\!\!\!  \;{\textstyle\sum\limits_{0\leq k\leq 2}}\! c^{2-k}\!
\displaystyle  \int_0^{t}\!  
\tpddt{\mathbf{Z}}_{\boldsymbol{\xi}}^{[k]}\big(t,t^{\mathrm{r}}\big)
(t- t^{\mathrm{r}})^{2-k} \drm{t^{\mathrm{r}}}  \biggl. \biggr].
\end{alignat}
 Here, $\boldsymbol{\xi}(t) \equiv (\qV,\vV,{\color{red}\aV})(t)$ 
and 
${\mathbf{Z}}_{\boldsymbol{\xi}}^{[2]}(t,t) :=\lim_{t^{\mathrm{r}}\to t}
{\mathbf{Z}}_{\boldsymbol{\xi}}^{[k]}\big(t,t^{\mathrm{r}}\big)$,
where
\begin{alignat}{1}\label{Zdef}
{\mathbf{Z}}_{\boldsymbol{\xi}}^{[k]}\big(t,t^{\mathrm{r}}\big) = 
 \displaystyle  \int_0^{2\pi}\!\! \int_0^{\pi}\!
\left(1-\tfrac1c{v^{}(t^\mathrm{r})}\cos\vartheta\right)
 \boldsymbol{\pi}_{\boldsymbol{\xi}}^{[k]}\big(t,\qV(t^\mathrm{r}) + c(t-t^\mathrm{r})\nV 
\big) 
\sin\vartheta \drm{\vartheta}\drm{\varphi}\,,
 \end{alignat}
with $v(t)$ defined by $v(t) |\EV^{\mbox{\tiny{hom}}}| \equiv\vV(t)\cdot\EV^{\mbox{\tiny{hom}}}$, and with
$\nV =\left(\sin\vartheta \cos\varphi,\;\sin\vartheta \sin\varphi ,\; \cos\vartheta \right)$
a normal vector to the retarded sphere of radius $r=c(t-t^\mathrm{r})$, where we measure 
$\vartheta$ from the $\EV^{\mbox{\tiny{hom}}}$ direction and $\varphi$ from an arbitrary axis $\perp\EV^{\mbox{\tiny{hom}}}$.

\newpage
 Moreover, the $\boldsymbol{\pi}_{\boldsymbol{\xi}}^{[k]}(t,\sV)$ with $k\in\{0,1,2\}$ and $\sV\neq\qV$ are defined as follows.
 We set
\begin{alignat}{1}
\mathrm{K}_{\boldsymbol{\xi}}(t',t,\sV) & := \label{rmK}
\tfrac{J_1\!\bigl(\varkappa\sqrt{c^2(t-t')^2-|\sV-\qV(t')|^2 }\bigr)}{\sqrt{c^2(t-t')^2-|\sV-\qV(t')|^2}^{\phantom{n}}},\\
\mathbf{K}_{\boldsymbol{\xi}}(t',t,\sV) & := \label{bfK}
\tfrac{J_2\!\bigl(\varkappa\sqrt{c^2(t-t')^2-|\sV-\qV(t')|^2 }\bigr)}{{c^2(t-t')^2-|\sV-\qV(t')|^2}^{\phantom{n}} }
 \left(\sV-\qV(t')- \vV(t')(t-t')\right), 
\end{alignat}
and note that 
\begin{alignat}{1}
\label{bfKxiNULL}
\int_{0}^{t^\mathrm{ret}_{\boldsymbol{\xi}^\circ}(t,\sV)} \!\!\!\!
 \mathbf{K}_{\boldsymbol{\xi}^\circ}(t',t,\sV)c\drm{t'} 
=
\int_0^{ct-|\sV|}\frac{J_2(\varkappa\sqrt{c^2(t-t')^2-|\sV|^2})}{c^2(t-t')^2-|\sV|^2}\sV\,d(ct'),
\end{alignat}
\begin{alignat}{1}
\label{rmKxiNULL}
 \int_{0}^{t^\mathrm{ret}_{\boldsymbol{\xi}^\circ}(t,\sV)}  \!\!\!\!\mathrm{K}_{\boldsymbol{\xi}^\circ}(t',t,\sV) c\drm{t'}
=
\int_0^{ct-|\sV|}\frac{J_1(\varkappa\sqrt{c^2(t-t')^2-|\sV|^2})}{\sqrt{c^2(t-t')^2-|\sV|^2}}\,d(ct').
\end{alignat}
 We will use $\big|_\mathrm{ret}$ to mean that $\qV(\tilde{t})$, $\vV(\tilde{t})$, ${\color{red}\aV}(\tilde{t})$ 
are evaluated at~$\tilde{t} = 
{t^\mathrm{ret}_{\boldsymbol{\xi}}}(t,\sV)$,
\textit{not} ${t^\mathrm{ret}_{\boldsymbol{\xi^\circ}}}(t,\sV)$.
  Then
\begin{alignat}{1}
\label{pi0}
 \boldsymbol{\pi}_{\boldsymbol{\xi}}^{[0]}(t,\sV) =
& - \varkappa^4 \frac14\left[
{\textstyle{
\frac{\left({\nV(\qV,\sV)} -\frac1c{\vV}\right)\crprd\color{black}\left(\frac1c{\vV}\crprd {\nV(\qV,\sV)} \right)}{
      \bigl({1-\frac1c {\vV}\cdot\nV(\qV,\sV)}\bigr)^{\!2} }
           }}\right]_{\mathrm{ret}}\\ \notag
&+ \varkappa^4\frac12\left[
{\textstyle{
\frac{ {\nV(\qV,\sV)}
-\frac1c{\vV}}{ {1-\frac1c {\vV}\cdot\nV(\qV,\sV)} }
             }}\right]_{\mathrm{ret}} 
\crprd \!
\int_{0}^{t^\mathrm{ret}_{\boldsymbol{\xi}}(t,\sV)}\!\!\!\!
{\vV(t')}\crprd \mathbf{K}_{\boldsymbol{\xi}}(t',t,\sV)\drm{t'} 
\\ \notag
& - \varkappa^4\frac12\left[{\textstyle{\frac{ \color{black} \frac1c{\vV}\crprd {\nV(\qV,\sV)} }{
      1-\frac1c {\vV}\cdot\nV(\qV,\sV)} }}\right]_{\mathrm{ret}} 
\crprd 
\int_{0}^{t^\mathrm{ret}_{\boldsymbol{\xi}}(t,\sV)}\!\!\!\!
 c\mathbf{K}_{\boldsymbol{\xi}}(t',t,\sV)\drm{t'}
\\  \notag
&
 +\varkappa^4\frac12\left[{\textstyle{\frac{ \color{black} \frac1c{\vV}\crprd {\nV(\qV,\sV)} }{
      1-\frac1c {\vV}\cdot\nV(\qV,\sV)} }}\right]_{\mathrm{ret}} 
\crprd 
\left(
 \tfrac{1-(1+\varkappa|\sV|)e^{-\varkappa|\sV|}}{\varkappa^2|\sV|^2}-\tfrac12\right)\tfrac{\sV}{|\sV|}
\\ \notag
&+\varkappa^4\frac12\left[{\textstyle{\frac{ \color{black} \frac1c{\vV}\crprd {\nV(\qV,\sV)} }{
      1-\frac1c {\vV}\cdot\nV(\qV,\sV)} }}\right]_{\mathrm{ret}} 
\crprd \int_{0}^{t^\mathrm{ret}_{\boldsymbol{\xi}^\circ}(t,\sV)} \!\!\!\! c\mathbf{K}_{\boldsymbol{\xi}^\circ}(t',t,\sV)\drm{t'} 
\\ \notag
& - \varkappa^4 \int_{0}^{t^\mathrm{ret}_{\boldsymbol{\xi}}(t,\sV)} \!\!\!\!
 c\mathbf{K}_{\boldsymbol{\xi}}(t',t,\sV)\drm{t'} \crprd \int_{0}^{t^\mathrm{ret}_{\boldsymbol{\xi}}(t,\sV)} \!\!\!\!
{\vV(t')}\crprd \mathbf{K}_{\boldsymbol{\xi}}(t',t,\sV)\drm{t'} 
\\ \notag
& +\varkappa^4 \left(
 \tfrac{1-(1+\varkappa|\sV|)e^{-\varkappa|\sV|}}{\varkappa^2|\sV|^2}-\tfrac12\right)\tfrac{\sV}{|\sV|}\crprd 
 \int_{0}^{t^\mathrm{ret}_{\boldsymbol{\xi}}(t,\sV)} \!\!\!\!
{\vV(t')}\crprd\mathbf{K}_{\boldsymbol{\xi}}(t',t,\sV)\drm{t'} 
\\ \notag
&+\varkappa^4
\int_{0}^{t^\mathrm{ret}_{\boldsymbol{\xi}^\circ}(t,\sV)} \!\!\!\! c\mathbf{K}_{\boldsymbol{\xi}^\circ}(t',t,\sV)\drm{t'} 
\crprd \int_{0}^{t^\mathrm{ret}_{\boldsymbol{\xi}}(t,\sV)} \!\!\!\!
{\vV(t')}\crprd \mathbf{K}_{\boldsymbol{\xi}}(t',t,\sV)\drm{t'} 
\\ \notag
& - \varkappa^4 c\int_{0}^{t^\mathrm{ret}_{\boldsymbol{\xi}}(t,\sV)} \!\!\!\! \mathrm{K}_{\boldsymbol{\xi}}(t',t,\sV)\drm{t'} 
 \int_{0}^{t^\mathrm{ret}_{\boldsymbol{\xi}}(t,\sV)}  \!\!\!\!\mathrm{K}_{\boldsymbol{\xi}}(t',t,\sV) {\vV}(t')\drm{t'}\,,
\\
\notag
& - \varkappa^3 \tfrac{1-e^{-\varkappa|\sV|}}{|\sV|}
 \int_{0}^{t^\mathrm{ret}_{\boldsymbol{\xi}}(t,\sV)}  \!\!\!\!\mathrm{K}_{\boldsymbol{\xi}}(t',t,\sV) {\vV}(t')\drm{t'}\,,
\\ \notag
& + \varkappa^4
 \int_{0}^{t^\mathrm{ret}_{\boldsymbol{\xi}^\circ}(t,\sV)}  \!\!\!\!\mathrm{K}_{\boldsymbol{\xi}^\circ}(t',t,\sV) c\drm{t'}
 \int_{0}^{t^\mathrm{ret}_{\boldsymbol{\xi}}(t,\sV)}  \!\!\!\!\mathrm{K}_{\boldsymbol{\xi}}(t',t,\sV) {\vV}(t')\drm{t'}\,,
\end{alignat}
and
\begin{alignat}{1}
\label{pi1}
 \boldsymbol{\pi}_{\boldsymbol{\xi}}^{[1]}(t,\sV) = 
& - \varkappa^2 
\left[
{\textstyle{
{\color{black}
{\nV(\qV,\sV)}\frac{\left({\nV(\qV,\sV)}\crprd [{ 
{\nV(\qV,\sV)} 
\crprd {\color{red}\aV} }]\right)\cdot\frac1c\vV}{
       c^2 \bigl({1-\frac1c {\vV}\cdot\nV(\qV,\sV)}\bigr)^{\!4} }
    } +
{\nV(\qV,\sV)} \crprd\frac{ {\nV(\qV,\sV)} 
\crprd {\color{red}\aV} }{
      2c^2 \bigl({1-\frac1c {\vV}\cdot\nV(\qV,\sV)}\bigr)^{\!3} }
}}\right]_{\mathrm{ret}}\\ \notag
&- \varkappa^2\left[
{\textstyle{
{\nV(\qV,\sV)}\crprd\frac{ {\nV(\qV,\sV)} 
\crprd {\color{red}\aV} }{
      c^2 \bigl({1-\frac1c {\vV}\cdot\nV(\qV,\sV)}\bigr)^{\!3} }
}}\right]_{\mathrm{ret}} \!\!
\crprd \!
\int_{0}^{t^\mathrm{ret}_{\boldsymbol{\xi}}(t,\sV)}\!\!\!\!
{\vV(t')}\crprd \mathbf{K}_{\boldsymbol{\xi}}(t',t,\sV)\drm{t'} 
\\ \notag
& + \varkappa^2\left[\nV(\qV,\sV)\crprd \biggl[{\textstyle{\nV(\qV,\sV)\crprd 
\frac{{\nV(\qV,\sV)}_{\phantom{!\!}} 
\crprd{\color{red}\aV} }{
      c^2\bigl({1-\frac1c {\vV}\cdot\nV(\qV,\sV)}\bigr)^{\!3} }
}}\biggr]\right]_{\mathrm{ret}} \!\!\! 
\crprd\! \int_{0}^{t^\mathrm{ret}_{\boldsymbol{\xi}}(t,\sV)} \!\!\!\!
 c\mathbf{K}_{\boldsymbol{\xi}}(t',t,\sV)\drm{t'} 
\\ \notag
& - \varkappa^2\left[\nV(\qV,\sV)\crprd \biggl[{\textstyle{\nV(\qV,\sV)\crprd 
\frac{{\nV(\qV,\sV)}_{\phantom{!\!}} 
\crprd{\color{red}\aV} }{
      c^2\bigl({1-\frac1c {\vV}\cdot\nV(\qV,\sV)}\bigr)^{\!3} }
}}\biggr]\right]_{\mathrm{ret}} \!\!\! 
\crprd\! \left(
 \tfrac{1-(1+\varkappa|\sV|)e^{-\varkappa|\sV|}}{\varkappa^2|\sV|^2}-\tfrac12\right)\tfrac{\sV}{|\sV|}\
\\ \notag
& - \varkappa^2\left[\nV(\qV,\sV)\crprd \biggl[{\textstyle{\nV(\qV,\sV)\crprd 
\frac{{\nV(\qV,\sV)}_{\phantom{!\!}} 
\crprd{\color{red}\aV} }{
      c^2\bigl({1-\frac1c {\vV}\cdot\nV(\qV,\sV)}\bigr)^{\!3} }
}}\biggr]\right]_{\mathrm{ret}} \!\!\! 
\crprd\! \int_{0}^{t^\mathrm{ret}_{\boldsymbol{\xi}^\circ}(t,\sV)} \!\!\!\! c\mathbf{K}_{\boldsymbol{\xi}^\circ}(t',t,\sV)\drm{t'} 
\\ \notag
& +\varkappa^3 
 \left[\textstyle\frac{1}{{1-\frac1c {\vV}\cdot\nV(\qV,\sV)} }\right]_{\mathrm{ret}} 
 \int_{0}^{t^\mathrm{ret}_{\boldsymbol{\xi}}(t,\sV)}\!\!\!\!
 \mathrm{K}_{\boldsymbol{\xi}}(t',t,\sV)\left[{\vV}({t^\mathrm{ret}_{\boldsymbol{\xi}}(t,\sV)})+{\vV}(t')\right]\drm{t'}\,,
\\ \notag
& +\varkappa^2 
 \left[\textstyle\frac{1}{{1-\frac1c {\vV}\cdot\nV(\qV,\sV)} }\right]_{\mathrm{ret}} 
\tfrac{1-e^{-\varkappa|\sV|}}{|\sV|}\tfrac1c{\vV}({t^\mathrm{ret}_{\boldsymbol{\xi}}(t,\sV)})
\\ \notag
& -\varkappa^3 
 \left[\textstyle\frac{1}{{1-\frac1c {\vV}\cdot\nV(\qV,\sV)} }\right]_{\mathrm{ret}} 
 \int_{0}^{t^\mathrm{ret}_{\boldsymbol{\xi}^\circ}(t,\sV)}  \!\!\!\!\mathrm{K}_{\boldsymbol{\xi}^\circ}(t',t,\sV) c\drm{t'}
\tfrac1c{\vV}({t^\mathrm{ret}_{\boldsymbol{\xi}}(t,\sV)}),
 \end{alignat}\vspace{-.5truecm}
and
\begin{alignat}{1}
\label{pi2}
\boldsymbol{\pi}_{\boldsymbol{\xi}}^{[2]}(t,s) = & - \varkappa^2
\left[\textstyle\frac{1}{\bigl({1-\frac1c {\vV}\cdot\nV(\qV,\sV)}\bigr)^{\!2} }\frac1c{\vV}
{\color{black} - \Big[\!{1-\tfrac{1}{c^2}\big|\vV\big|^2}\!\Big]
\frac{ \left({\nV(\qV,\sV)}-\frac1c{\vV}\right) \crprd \left(\frac1c\vV\crprd \nV(\qV,\sV)\right) }{
      \bigl({1-\frac1c {\vV}\cdot\nV(\qV,\sV)}\bigr)^4 }}
\right]_{\mathrm{ret}} 
\\ \notag
&  +\varkappa^2 \left[\Big[\!{1-\tfrac{1}{c^2}\big|\vV\big|^2}\!\Big] {\textstyle{
\frac{ \frac1c{\vV}\crprd {\nV(\qV,\sV)}_{\phantom{!\!}} }{
      \bigl({1-\frac1c {\vV}\cdot\nV(\qV,\sV)}\bigr)^{\!3} }
}}\right]_{\mathrm{ret}} \crprd\int_{0}^{t^\mathrm{ret}_{\boldsymbol{\xi}}(t,\sV)} \!\!\!\!
 c\mathbf{K}_{\boldsymbol{\xi}}(t',t,\sV)\drm{t'} \\
\notag
&  - \varkappa^2\left[\Big[\!{1-\tfrac{1}{c^2}\big|\vV\big|^2}\!\Big] {\textstyle{
\frac{ \frac1c{\vV}\crprd {\nV(\qV,\sV)}_{\phantom{!\!}} }{
      \bigl({1-\frac1c {\vV}\cdot\nV(\qV,\sV)}\bigr)^{\!3} }
}}\right]_{\mathrm{ret}} \crprd\left(
 \tfrac{1-(1+\varkappa|\sV|)e^{-\varkappa|\sV|}}{\varkappa^2|\sV|^2}-\tfrac12\right)\tfrac{\sV}{|\sV|}
\\
\notag  
&  -\varkappa^2 \left[\Big[\!{1-\tfrac{1}{c^2}\big|\vV\big|^2}\!\Big]{\textstyle{
\frac{\frac1c{\vV}\crprd  {\nV(\qV,\sV)}_{\phantom{!\!}} }{
      \bigl({1-\frac1c {\vV}\cdot\nV(\qV,\sV)}\bigr)^{\!3} }
}}\right]_{\mathrm{ret}} \crprd \int_{0}^{t^\mathrm{ret}_{\boldsymbol{\xi}^\circ}(t,\sV)} \!\!\!\!
 c\mathbf{K}_{\boldsymbol{\xi}^\circ}(t',t,\sV)\drm{t'} 
 \\
\notag
& -  \varkappa^2\left[\Big[\!{1-\tfrac{1}{c^2}\big|\vV\big|^2}\!\Big]
{\textstyle{
\frac{ {\nV(\qV,\sV)}_{\phantom{!\!}}-\frac1c{\vV} }{
      \bigl({1-\frac1c {\vV}\cdot\nV(\qV,\sV)}\bigr)^{\!3} }
}}\right]_{\mathrm{ret}} 
\crprd \!
\int_{0}^{t^\mathrm{ret}_{\boldsymbol{\xi}}(t,\sV)} \!\!\!\!
{\vV(t')}\crprd \mathbf{K}_{\boldsymbol{\xi}}(t',t,\sV)\drm{t'} .
\end{alignat}

 Although this is an intimidating list of integrals, we can already extract an important
conclusion: The equation of motion for our point charge does not feature time-derivatives 
of the particle position $\qV(t)$ higher than second order. 
 This result holds also for BLTP electrodynamics in general \cite{KiePRD}.
 Hence, BLTP electrodynamics with point charges does not suffer from the $\dddot\qV(t)$ problem.
                \subsection{The small-$\varkappa$ regime}\label{sec:MBLTPkappaSMALL}
 To make further progress in the evaluation of the integrals we will concentrate our
efforts on an asymptotic analysis of the small-$\varkappa$ regime.
 We make a formal power series expansion about $\varkappa=0$ given by
 $\fV^{\mbox{\tiny{self}}}[\qV,\vV;{\color{red}\aV}](t) = \sum_{n=0}^\infty \FV^{(n)}_0[\qV,\vV;{\color{red}\aV}](t)$,
where $\FV^{(n)}_0 \propto \varkappa^n$;
 the subscript ${}_0$ at $\FV^{(n)}_0$ indicates that we are expanding about $\varkappa=0$. 
 It is manifest that the terms $O(\varkappa^0)$ and $O(\varkappa^1)$ vanish identically, so we need to discuss terms
$O(\varkappa^n)$ for $n\geq 2$.
 Several of the spherical integrations can been carried out explicitly in terms of well-known functions.
 In particular, the contributions $\FV^{(2)}_0$ and $\FV^{(3)}_0$ can be computed explicitly.

                \subsubsection{Radiation-reaction at $O(\varkappa^2)$}\label{sec:MBLTPkappaSQR}
\vspace{-7pt}
 
 To arrive at the $O(\varkappa^2)$ contribution we
divide the expressions for $\boldsymbol{\pi}^{[k]}_{\boldsymbol{\xi}}$ by $\varkappa^2$ and
take $\varkappa\to 0$. 
 The only two terms that survive in the limit are those in the first line of rhs\refeq{pi1} and rhs\refeq{pi2},
respectively (indicated below by a superscript ${}^{,1}$; later, also superscripts ${}^{,3}$,
${}^{,4}$, ${}^{,7}$ will appear).
 Carrying out the pertinent integrations in \refeq{Zdef} one notes that the result only depends on $t^{\mathrm{r}}$, 
not on $t$, so that the third line of rhs\refeq{eq:selfFexpl} vanishes at $O(\varkappa^2)$.
 Thus,
\begin{alignat}{1}
\label{eq:SELFforceATorderTWO}
\tfrac{4\pi } {\ee^2} \FV^{(2)}_0(t) = \;
 - {\mathbf{Z}}_{\boldsymbol{\xi}}^{[2],1}(t,t) 
- c \displaystyle  \int_0^{t}\! {\mathbf{Z}}_{\boldsymbol{\xi}}^{[1],1}\big(t,t^{\mbox{\tiny{r}}}\big)\drm{t^{\mbox{\tiny{r}}}} .
\end{alignat}
 Explicitly, \refeq{eq:SELFforceATorderTWO} reads
\begin{alignat}{1}
\label{eq:SELFforceATorderTWOeval}
\hspace{-1truecm} \FV^{(2)}_0(t) = 
& 
- \frac12 \eEL^2 \varkappa^2 \frac{\vV(t)}{{v(t)}} 
\left[ 2 \frac{c}{v(t)} - \frac{c^2}{v(t)^2} \ln \frac{1+\frac1c{v(t)}}{1-\frac1c{v(t)}} \right] 
\\ \notag
& + \eEL^2 \varkappa^2  \int_0^{t}\! 
\frac{c^3}{v(t^{\mbox{\tiny{r}}})^3} \Biggl[ 
 \tfrac1c{v(t^{\mbox{\tiny{r}}})} \frac{2 -  \frac{v(t^{\mbox{\tiny{r}}})^2}{c^2}}{1-\frac{v(t^{\mbox{\tiny{r}}})^2}{c^2}}  - 
\ln\frac{1+\frac1c v(t^{\mbox{\tiny{r}}})}{1-\frac1c v(t^{\mbox{\tiny{r}}})}
\Biggr] \biggr.\tfrac{1}{c} {\color{red}\aV}(t^{\mbox{\tiny{r}}}) \drm{t^{\mbox{\tiny{r}}}}
\end{alignat}
 The term in the first line of rhs\refeq{eq:SELFforceATorderTWOeval} is the contribution from the first line
of rhs\refeq{pi2}, the term in the second line of rhs\refeq{eq:SELFforceATorderTWOeval} is the contribution 
from the first line of rhs\refeq{pi1}. 
 Since for straight-line motion $\vV(t)$ and ${\color{red}\aV}(t)$ are collinear, and ${\color{red}\aV}(t)=\dot\vV(t)$, 
one can carry out the time 
integration in the second line of rhs\refeq{eq:SELFforceATorderTWOeval} in terms of elementary functions of $\vV/c$, and 
a few algebraic manipulations then give (see the erratum in \cite{KiePRD})
\begin{alignat}{1}
 \FV^{(2)}_0(t) = \label{eq:SELFforceATorderTWOmerged}
 \, \NullV.
\end{alignat}
 In this problem of straight line motion in a constant external electric field, with the particle starting from rest,
the BLTP radiation-reaction force vanishes \emph{exactly} at $O(\varkappa^2)$.

                \subsubsection{Radiation-reaction at $O(\varkappa^3)$}\label{sec:MBLTPkappaTHREE}
\vspace{-7pt}

 We next evaluate the $O(\varkappa^3)$ contribution to the  radiation-reaction force for small $\varkappa$.
 To arrive at the $O(\varkappa^3)$ contribution, subtract the $O(\varkappa^2)$ terms
from the expressions for $\boldsymbol{\pi}^{[k]}_{\boldsymbol{\xi}}$, divide the result
by $\varkappa^3$ and take the limit $\varkappa\to 0$. 
 This yields the contributions from three $\boldsymbol{\pi}^{[k]}_{\boldsymbol{\xi}}$ terms $\propto\varkappa^3$, namely 
$\boldsymbol{\pi}^{[1],4}_{\boldsymbol{\xi}}$, $\boldsymbol{\pi}^{[1],7}_{\boldsymbol{\xi}}$, 
and $\boldsymbol{\pi}^{[2],3}_{\boldsymbol{\xi}}$.
 They contribute the following force $\propto\varkappa^3$, 
\begin{alignat}{1}
\label{eq:SELFforceEXPLICITexpandATorderTHREE}
\hspace{-1truecm} \FV^{(3)}_0(t) = 
& 
-  \eEL^2 \varkappa^3\qV(t) \\ \notag
&
- \eEL^2 \varkappa^3
 \tfrac{\color{black}{4}}{3}
\int_0^t\! \frac{c^2}{v(t^{\mbox{\tiny{r}}})^2} 
  \biggl\{ 1 +\tfrac12 \frac{c}{v(t^{\mbox{\tiny{r}}})}
\ln \frac{1-\frac1c{v(t^{\mbox{\tiny{r}}})}}{1+\frac1c{v(t^{\mbox{\tiny{r}}})}}  \biggr\} 
 (t-t^{\mbox{\tiny{r}}}) {\color{red}\aV}(t^{\mbox{\tiny{r}}})  \drm{t^{\mbox{\tiny{r}}}} \\ \notag
& 
{\color{black}-} \eEL^2 \varkappa^3
 \tfrac{2}{3}
 \int_0^t\! \left({1-\frac{v(t^{\mbox{\tiny{r}}})^2}{c^2}}\right)\frac{c}{v(t^{\mbox{\tiny{r}}})} 
  \biggl\{ 1 +\tfrac12 \frac{c}{v(t^{\mbox{\tiny{r}}})}
\ln \frac{1-\frac1c{v(t^{\mbox{\tiny{r}}})}}
{1+\frac1c{v(t^{\mbox{\tiny{r}}})}}  \biggr\} 
c \drm{t^{\mbox{\tiny{r}}}} \tfrac{\EV^{\mbox{\tiny{hom}}}}{|\EV^{\mbox{\tiny{hom}}}|}
\end{alignat}
 Integration by parts yields for the integral in the second line of rhs\refeq{eq:SELFforceEXPLICITexpandATorderTHREE} 
\begin{alignat}{1}
\label{eq:OthreeINTbyPARTS}
&
\int_0^t\! \frac{c^2}{v(t^{\mbox{\tiny{r}}})^2} 
  \biggl\{ 1 +\tfrac12 \frac{c}{v(t^{\mbox{\tiny{r}}})}
\ln \frac{1-\frac1c{v(t^{\mbox{\tiny{r}}})}}{1+\frac1c{v(t^{\mbox{\tiny{r}}})}}  \biggr\} 
 (t-t^{\mbox{\tiny{r}}}) {\color{red}\aV}(t^{\mbox{\tiny{r}}})  \drm{t^{\mbox{\tiny{r}}}} = \\ \label{eq:OthreeINTbyPARTSb}
& 
\int_0^t\! c\int_0^{\vV(t^{\mbox{\tiny{r}}})/c} \frac{1}{x^2} 
  \biggl\{ 1 +\tfrac12 \frac{1}{x}
\ln \frac{1-x}{1+x}  \biggr\} 
 \drm{x}
 \drm{t^{\mbox{\tiny{r}}}}  = \\ \notag
& 
-\frac12\int_0^t\! \frac{c}{v(t^{\mbox{\tiny{r}}})} \left[1+\left(1-\frac{v(t^{\mbox{\tiny{r}}})^2}{c^2}\right)
\tfrac12 \frac{c}{v(t^{\mbox{\tiny{r}}})}
\ln \frac{1-\frac1c{v(t^{\mbox{\tiny{r}}})}}{1+\frac1c{v(t^{\mbox{\tiny{r}}})}}  \right]
c \drm{t^{\mbox{\tiny{r}}}}  
\end{alignat}
 Comparison with the third line of rhs\refeq{eq:SELFforceEXPLICITexpandATorderTHREE} reveals cancelations, 
and we end up with
\begin{alignat}{1}
\label{eq:SELFforceEXPLICITexpandATorderTHREEfinal}
 \FV^{(3)}_0(t) = 
  - \frac13  \eEL^2 \varkappa^3\qV(t) .
\end{alignat}

 This is a very surprising result: the $O(\varkappa^3)$ term of the radiation-reaction force in our initial value problem 
is a harmonic oscillator force!
 This result relies on the particular setup of the initial data and the geometry of the problem, but not more.

     \section{The Volterra equation for the acceleration}  

 The equation of motion can be recast as a Volterra integral equation for the acceleration, 
\begin{equation}\label{Volterra}
{\color{red}\aV}(t)= W[\vV]\cdot\left(\eEL \EV^{\mbox{\tiny{hom}}} + \fV^{\mbox{\tiny{self}}}[\qV,\vV;{\color{red}\aV}]\right)(t).
\end{equation}
 Here,
\begin{equation}
 W[\vV]
:= \label{eq:FtoAmapINv}
\textstyle
\frac{1}{\mbare}\sqrt{1-\frac{|\vV|^2}{c^2}}
\left[\ID - \frac{1}{c^2}\vV\otimes \vV\right],
\end{equation}
which for motion along $\EV^{\mbox{\tiny{hom}}}$ is the same as
\begin{equation}
 W[\vV]
:= \label{eq:FtoAmapINvAGAIN}
\textstyle
\frac{1}{\mbare}\left(1-\frac{|\vV|^2}{c^2}\right)^{3/2}\ID .
\end{equation}
 In \cite{KTZonBLTP} we show that the Volterra equation can be uniquely solved to yield ${\color{red}\aV}$ as a nonlinear
 expression in $\qV$ and $\vV$, posing a second-order initial value problem for $\qV(t)$.

     \subsection{The Volterra equation to $O(\varkappa^3)$ (small $\varkappa$)} 

 With the radiation-reaction force evaluated to $O(\varkappa^3)$ we obtain the equation of motion 
\begin{equation}\label{VolterraATorderTHREE}
{\color{red}\aV}(t)= 
\textstyle
\frac{1}{\mbare}\left(1-\frac{1}{c^2}|\vV(t)|^2\right)^{3/2}\left(\eEL \EV^{\mbox{\tiny{hom}}} -\frac13  \eEL^2 \varkappa^3\qV(t) \right).
\end{equation}
 Eq.(\ref{VolterraATorderTHREE}) is equivalent to the problem of special-relativistic test particle motion in a 
harmonic oscillator potential, featuring time-periodic solutions with conserved particle energy 
\begin{equation}\label{EFFECTIVEenergy}
U = \sqrt{\mbare^2 c^4 + |\pV|^2 c^2} - \eEL  \EV^{\mbox{\tiny{hom}}}\cdot\qV  +\tfrac16  \eEL^2 \varkappa^3|\qV|^2.
\end{equation}

 It is not clear to us whether this means that the validity of the $O(\varkappa^3)$ approximation 
is restricted to short times $\varkappa ct\ll 1$ (see Fig.1, which looks reasonable) or whether such
periodic motion over longer times is a genuine feature of BLTP electrodynamics as long as $\varkappa e^2/\mbare c^2\ll 1$. 
 In the latter case BLTP electrodynamics would presumably be eliminated for good from the list of contenders for a realistic
classical electrodynamics.
\vspace{-.3truecm}
\begin{figure}[H]
  \includegraphics[width = 6truecm,scale=2]{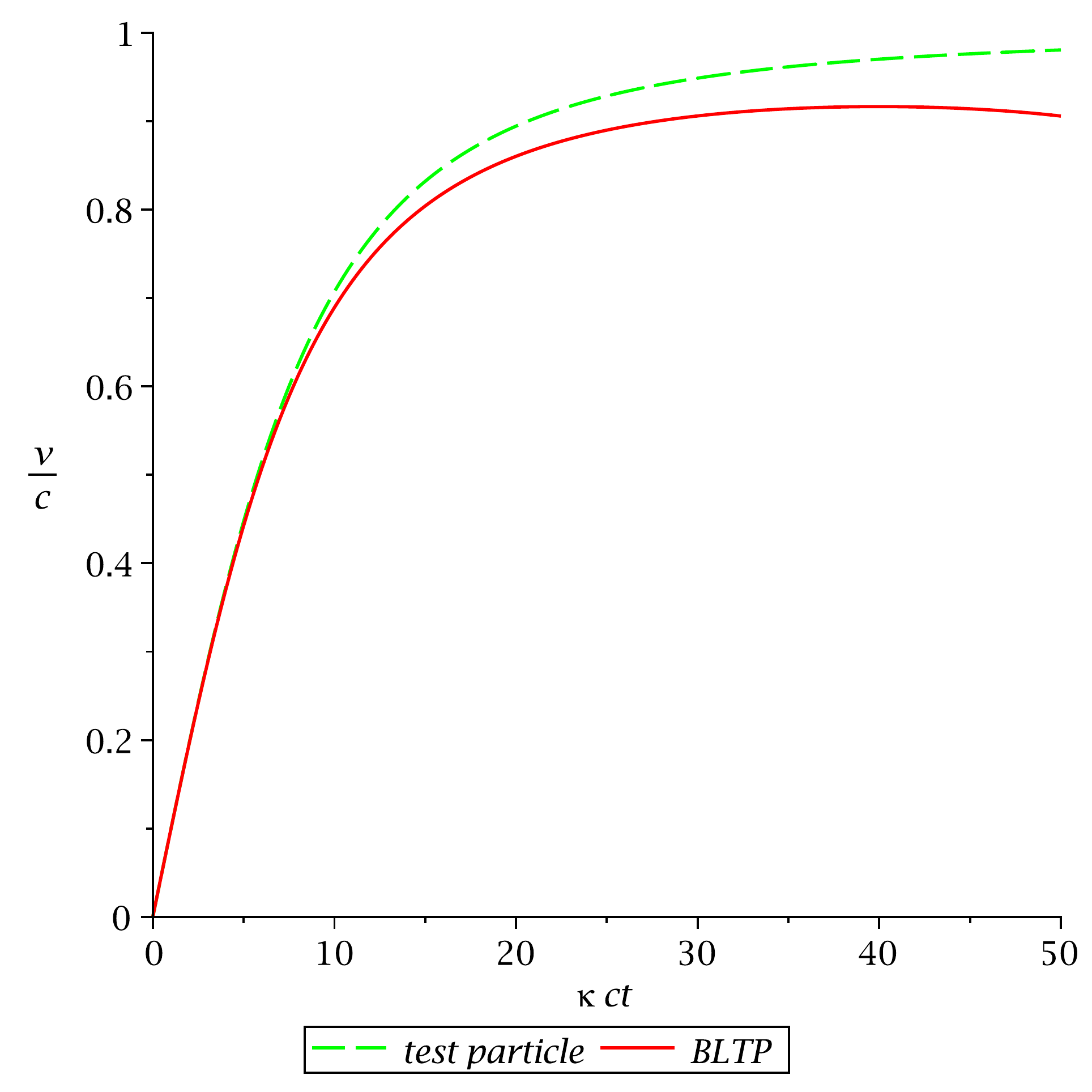} \vspace{-.3truecm}
\caption{\footnotesize{The velocity of a point charge, starting from rest in a constant applied electrostatic 
field $\EV^{\mbox{\tiny{hom}}}=10e\varkappa^2$, vs. time, as per test particle theory (dashed curve), resp. BLTP electrodynamics 
with radiation-reaction included to $O(\varkappa^3)$ (continuous curve), when $\varkappa e^2/\mbare c^2 =0.01$.
 The period of the velocity of the BLTP motion is $\varkappa c T = 160$. The test particle's velocity asymptotes to $c$.}
}
\end{figure}

\vspace{-.7truecm}
     \section{Summary and Outlook}

We have shown that BLTP electrodynamics, as defined in \cite{KiePRD}, accounts for radiation-reaction effects on the point
charge motion along a constant applied electric field, thereby passing a litmus test that other models (in particular, 
the Landau--Lifshitz and Eliezer--Ford--O'Connell equations of motion) fail. 
 Our results are based on a small-$\varkappa$ expansion of the BLTP force expression. 
 This is acceptable for our ``proof-of-concept'' demonstration.
 However, BLTP electrodynamics is physically viable at most in the large $\varkappa$ regime \cite{CKP} --- if at all.
 An assessment of the large-$\varkappa$ regime we leave to some future work. 

\newpage
\noindent 
\textbf{Acknowledgement:} We thank the referee for interesting comments that have prompted us to improve the presentation.


\begin{thebibliography}{[9999999]}\footnotesize{
\bibitem[ApKi2001]{AppKieAOP}
\vskip-5pt
        Appel, W., and Kiessling, M. K.-H.,
        {\it Mass and Spin Renormalization in Lorentz Electrodynamics},
        Annals Phys. \textbf{289}, 24--83 (2001).
\bibitem[Bop1940]{BoppA}
\vskip-5pt
        Bopp, F.,
        \textit{Eine lineare Theorie des Elektrons},
        Annalen Phys. \textbf{430}, 345--384 (1940).
\bibitem[Bop1943]{BoppB}
\vskip-5pt
        Bopp, F.,
        \textit{Lineare Theorie des Elektrons. II}, Annalen Phys. \textbf{434}, 573--608 (1943).
\bibitem[BoIn1934]{BornInfeldBb} 
\vskip-7pt
	Born, M., and Infeld, L.,
	\textit{Foundation of the new field theory},
	Proc. Roy. Soc. London \textbf{A 144}, 425--451 (1934).
\bibitem[CKP2019]{CKP}
\vskip-7pt
        Carley, H.K., Kiessling, M.K.-H., and Perlick, V.,
        \textit{On the Schr\"odinger spectrum of a Hydrogen atom with electrostatic Bopp--Land\'e--Thomas--Podolsky 
          interaction between electron and proton}, 
        Int. J. Mod. Phys. A \textbf{34}, 1950146 (23pp.) (2019).

\bibitem[DeHa2016]{DeckertHartenstein}
\vskip-5pt
	Deckert, D.-A., 
        and
        Hartenstein, V.,     
                \textit{On the initial value formulation of
		   classical electrodynamics}
	        J. Phys. A: Math. Theor. \textbf{49},  445202 (19pp.) (2016).
\bibitem[GHW2009]{GHW} 
\vskip-.3truecm
        Gralla, S.E.,
        Harte, A.,
        and
        Wald, R.M.,
        \textit{A Rigorous Derivation of Electromagnetic Self-Force},
        Phys. Rev. D \textbf{80}:024031 (2009).
\bibitem[GPT2015]{GratusETal}
\vskip-.3truecm
	Gratus, J., Perlick, V., and Tucker, R.W.,
        \textit{On the self-force in Bopp--Podolsky electrodynamics},
        J. Phys. A: Math. Theor. \textbf{48}, 435401 (28pp.) (2015).
\bibitem[Hetal2021]{VuMaria}
\vskip-.3truecm
	Hoang, V., Radosz, M., Harb, A., DeLeon, A.,  and Baza, A.,
        \textit{Radiation reaction in higher-order electrodynamics},
        J. Math. Phys. \textbf{62}, 072901 (31pp.) (2021).
\bibitem[Jac1975]{JacksonBOOKb}  
\vskip-7pt
        Jackson, J.D.,   
                \textit{Classical electrodynamics}, 
       	J. Wiley \& Sons, New York 
	$2^{nd}$ ed. (1975).
\bibitem[Kie2019]{KiePRD}
\vskip-5pt
	Kiessling, M.K.-H.,
        \textit{Force on a point charge source of the classical electromagnetic field},
        Phys. Rev. D \textbf{100}, 065012 (2019);
        \textit{Erratum} ibid. \textbf{101}, 109901(E) (2020).
\bibitem[KTZ2023]{KTZonBLTP}
\vskip-7pt
	Kiessling, M.K.-H.,
        and
        Tahvildar-Zadeh, A. S.,
        \textit{Bopp-Land\'e-Thomas-Podolsky electrodynamics as initial value problem},
        in preparation (2023).
\bibitem[Lan1941]{Lande}
\vskip-5pt
	Land\'e, A., 
        \textit{Finite Self-Energies in Radiation Theory. Part I},
        Phys. Rev. \textbf{60}, 121--126 (1941).
\bibitem[LaTh1941]{LandeThomas}
\vskip-5pt
	Land\'e, A., and Thomas, L.H.,
        \textit{Finite Self-Energies in Radiation Theory. Part II},
        Phys. Rev. \textbf{60}, 514--523 (1941).
\bibitem[Lau1909]{Laue}
\vskip-5pt
	v. Laue, M., 
        \textit{Die Wellenstrahlung einer bewegten Punktladung nach dem Relativit\"ats\-prinzip},
        Annalen Phys. \textbf{28}, 436--442 (1909).
\bibitem[Lor1904]{LorentzENCYCLOP}
\vskip-5pt
        Lorentz, H.A., 
                {\it Weiterbildung der Maxwell'schen Theorie: Elektronentheorie.},
        Encyklop\"adie d. Mathematischen Wissenschaften ${\bf V}2$,
        Art. 14, pp. 145--288 (1904).
\bibitem[Mil1998]{MillerBOOK} 
\vskip-5pt
        Miller, A. I., 
                {\sl Albert Einstein's special theory of relativity},
        Springer, New York (1998).
\bibitem[PMD2006]{PMD}
\vskip-5pt
        de Parga, A., Mares, R., and Dominguez, S.,
        {\it An unphysical result for the Landau--Lifshitz equation of motion for a charged particle},
        Rev. Mex. Fis. {\bf 52}, 139--142 (2006).
\bibitem[Pod1942]{Podolsky}
\vskip-5pt
        Podolsky, B.,
        \textit{A generalized electrodynamics. Part I: Non-quantum},
        Phys. Rev. \textbf{62}, 68--71 (1942).
\bibitem[PPV2011]{PoissonETal}
\vskip-5pt
   Poisson, E., Pound, A., and Vega, I.,
   \emph{The motion of point particles in curved spacetime},
 Living Rev. Rel. \textbf{14},7(190) (2011).
\bibitem[Spo2004]{Spohn}
\vskip-7pt
       Spohn, H.,
       \textit{Dynamics of charged particles and their radiation fields},
       Cambridge UP (2004).
\bibitem[Zay2014]{Zayats}
\vskip-5pt
        Zayats, A.E.,
        \textit{Self-interaction in the Bopp-Podolsky electrodynamics:
          Can the observable mass of a charged particle depend on its acceleration?},
        Annals Phys. (NY)\textbf{342}, 11--20 (2014).
}
\end{thebibliography}
\end{document}